\documentclass[12pt]{iopart}
\usepackage{graphicx}
\usepackage{dcolumn}
\usepackage{bm}

\usepackage{graphicx}     
\usepackage{epsf}

\begin{document}

\title[Symmetry Breaking in Two-Channel Exclusion Processes]{Spontaneous Symmetry Breaking in  Two-Channel Asymmetric Exclusion Processes with Narrow Entrances}
\author{Ekaterina Pronina and  Anatoly B. Kolomeisky} 
\address{Department of Chemistry, Rice University, Houston, TX 77005-1892}

\begin{abstract}

Multi-particle non-equilibrium dynamics in two-channel asymmetric exclusion processes with narrow entrances is investigated theoretically. Particles move on two parallel lattices in opposite directions without changing them, while the channels are coupled only at the boundaries. A particle cannot enter the corresponding lane if the exit site of the other lane is occupied. Stationary phase diagrams, particle currents and densities are calculated in a mean-field approximation. It is shown that there are four stationary phases in the system, with two of them exhibiting spontaneous symmetry breaking phenomena.  Extensive Monte Carlo computer simulations confirm qualitatively our predictions, although the phase boundaries and stationary properties deviate from the mean-field results. Computer simulations indicate that several dynamic and phase properties of the system have a strong size dependency, and  one of the stationary phases predicted by the mean-field theory disappears in the thermodynamic limit.

\end{abstract}


\ead{tolya@rice.edu}

\submitto{\JPA}

\maketitle

\section{Introduction}

Asymmetric simple exclusion processes (ASEP) have become an important tool for  understanding complex  non-equilibrium phenomena in chemistry, physics and biology \cite{derrida98,schutz}. In the absence of fundamental framework for non-equilibrium systems ASEP have emerged as reference models, and they play the role similar to the Ising model in the studies of equilibrium systems \cite{schutz03}.  ASEPs have been applied successfully to analyze the kinetics of biopolymerization \cite{macdonald68}, protein synthesis \cite{shaw03,kolomeisky04,chou04}, molecular transport through nanopores and channels \cite{chou99}, the motion of motor proteins along cytoskeleton filaments \cite{klumpp05,parmeggiani03,nishinari05}, and investigation  of car traffic processes \cite{helbing01,chowdhury00}.

ASEPs are one-dimensional  models where  particles, that interact via an exclusion potential, hop  along  discrete lattices.  Many non-equilibrium processes can be described by single-channel exclusion processes.  However, the necessity to analyze more realistic complex phenomena, that involve the transport along the parallel channels and existence of internal states, e.g.,  motor proteins, vehicular traffic and hopping of quantum dots, stimulated the development of multi-channel ASEPs \cite{popkov01,pronina04,pronina05,mitsudo05,harris05,pronina06,reichenbach06}. Theoretical analysis of multi-channel exclusion processes indicates that a coupling between channels has a strong effect on stationary-state phase diagrams and it leads to several unusual phenomena, such as localized domain walls \cite{pronina06,reichenbach06}.

One of the most intriguing phenomena observed in ASEPs is symmetry breaking when the microscopic symmetric dynamic rules lead to the existence of macroscopic asymmetric stationary-state properties for some sets of parameters \cite{popkov01,evans95,EFGM,arndt98,clincy01,popkov04,levine04,erickson05,willmann05}. However,  mechanisms  of symmetry breaking phenomena in ASEPs are still not well understood \cite{popkov04}. Currently, these phenomena are mostly described by mean-field approaches that neglect the correlations in the system, however, the agreement with computer simulations results is rather qualitative \cite{evans95,EFGM,arndt98,clincy01}.  There are suggestions that  dynamic amplifications of fluctuations might lead to a broken symmetry in the asymmetric exclusion processes \cite{willmann05}. It is also argued that the origin of this phenomena might be due to the effective boundary defects \cite{popkov04}. Note also that symmetry-breaking has not been obtained in the exact solutions of ASEPs \cite{EFGM,kolomeisky97}. 

Originally, the symmetry breaking has been observed in the single-lane exclusion process with two particles moving in opposite directions \cite{evans95,EFGM}. Using a mean-field approach and Monte Carlo computer simulations, it was shown that two stationary phases with broken symmetry  are possible for some part of the parameter space, although, surprisingly, the agreement between theoretical predictions and numerical calculations was not good, in contrast to other steady-state properties of other ASEPs where the mean-field description is rather successful \cite{derrida98,schutz}. There are controversial reports about the number of phases with broken symmetry. The existence of one of the symmetry-broken phase has been disputed  on the basis of computer simulations by Arndt et al. \cite{arndt98}. More extensive computer simulations \cite{clincy01} confirmed the presence of two asymmetric phases, but also it was argued that phase transitions to the symmetric phase is different from the mean-field predictions. However, extensive high-precision Monte Carlo data \cite{erickson05} indicated that the disputed intermediate phase is a finite-size effect and it disappears for very large systems. It is not clear if this observation of single asymmetric phase is correct for all ASEPs where the symmetry breaking have been observed. In addition, in single-channel two-species ASEP, where the symmetry breaking is observed, particles of different types interact with each other at every site of the lattice, i.e., the interaction is global. It is not clear how the interaction between these particles localized in the specific parts of the system  will affect the symmetry breaking.

In this paper  we investigate two-lane ASEPs with two types of particles moving along different channels in opposite directions. The particles cannot jump between the lanes, and two channels interact with each other only at the boundaries. the problem is motivated by the cellular transport of kinesin and dynein motor proteins that move along microtubules in opposite directions \cite{howard}. The system is analyzed using a mean-field theoretical approach and extensive Monte Carlo computer simulations at different system sizes, and we conclude that  the symmetry breaking is taking place in this system for some range of parameters. This observation supports the idea that symmetry breaking can be viewed as an effective boundary-induced phenomenon.

The paper is organized as follows. In section 2 we give a detailed description of the model and we solve it in the mean-field approximation. In section 3 we present and discuss Monte Carlo computer simulations, and we compare them with theoretical predictions. Finally, we summarize  and conclude in section 4.

\section{The Model and Theoretical Description} 

\subsection{Model}

We consider a system  of two parallel one-dimensional lattices with two types of particles moving in the different lanes in the opposite directions, as shown in Fig. 1. Two channels are identical and both have $L$ sites. The particles are interacting with hard-core exclusion potential which means that each lattice site can be empty, or it can be occupied by no more than one particle. The hoppings between the channels are not allowed, and the particle in the channel 1 (2) moves to the right (left) with the rate 1 if the next site is empty. The particles  can enter the corresponding channel with the rate $0<\alpha \le 1$ if the first site at the given lane is empty {\it and} the site $L$ at the other channel is unoccupied: see Fig. 1. The particles  exit both lanes with the same rate $0 < \beta \le 1$, that does not depend on the occupancy of the site 1 in the other channel.  Note that the particle at site $L-1$ can proceed to the  site $L$ if the exit site is unoccupied independently of the status of site 1 at the other lattice. Thus this model can be viewed as two-channel ASEP with narrow entrances.

\begin{figure}[h] 
\centering
\includegraphics[scale=0.7,clip=true]{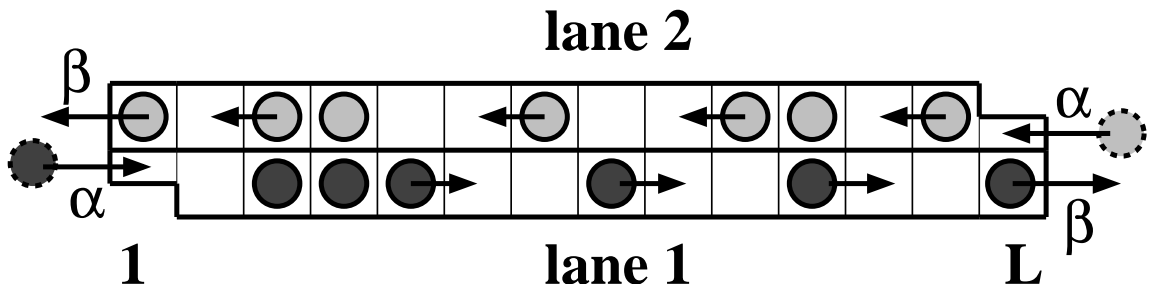}
\caption{Schematic view of two-lane ASEP with narrow entrances. Dark-gray particles move only along the channel 1, while light-gray particles can be found only on the channel 2. Allowed transitions are shown by arrows. Inter-channel transitions are not allowed. Entrance rates at both channels are equal to $\alpha$ if there are no particle at the exit of the other lane. Exit rates are equal to $\beta$ at both channels independently of the occupation status of the exit site  at the other channel.}\label{model}
\end{figure}

The model, as described above and illustrated in Fig. 1, is similar to a single-channel ASEP with two species where symmetry breaking has been previously observed \cite{evans95,EFGM,arndt98,clincy01,erickson05}. However, there are several important differences between two systems: in our model there are no restrictions on the bulk densities of particles in both channels and only the entrance to the channel is influenced by the other channel. The interaction between two types of particles is local in our system, in contrast to single-channel two-species ASEPs where particles interact along the whole lattice.

\subsection{Mean-Field Theory} \label{theory_intro}

Theoretical studies of ASEPs indicate that an approximate mean-field analysis provides a very good description of these processes, as confirmed by comparison with available exact solutions and computer simulations \cite{derrida98,schutz}. Following this approach, we introduce $p_{i}$ and $m_{i}$ as occupation variables for  the channel 1 and 2, respectively, such that  $p_{i}=1$ and/or $m_{i}=1$ if the corresponding site $i$ in the channel 1 and/or channel 2 is occupied, and 0 otherwise. The mean-field theory then assumes that there are no correlations in the probability of finding two particles at any two sites, i.e.,
\begin{equation}
\langle p_{i} p_{j} \rangle = \langle p_{i} \rangle \langle p_{j} \rangle = p_{i} p_{j}, \quad \langle m_{i} m_{j} \rangle = \langle m_{i} \rangle \langle m_{j} \rangle = m_{i} m_{j}.
\end{equation}

Our system with particles moving on two channels in opposite directions can be viewed as two single-lane ASEPs with exit rates $\beta$ and  entrance rates $\alpha_{1}$ and $\alpha_{2}$ for the lane 1 and 2, respectively. From the dynamic rules of the system we conclude that the effective entrance rates are given by 
\begin{equation} \label{eq_1}
\alpha_{1}=\alpha(1-m_{1}), \quad \alpha_{2}=\alpha(1-p_{L}).
\end{equation}
Large-time dynamics and  properties of single-channel ASEP are known exactly \cite{derrida98,schutz}. In this case there are three stationary phases, specified by the processes at the entrance, at the exit and in the bulk of the system.   For $\alpha < \beta$ and $\alpha < 1/2$  the entrance is the rate-limiting step in the overall dynamics, and the system is found in a low-density (LD) phase with the particle current and bulk density 
\begin{equation} \label{LD}
J_{LD}= \alpha(1- \alpha), \quad \rho_{bulk,LD}=\alpha.
\end{equation}
When the exit is the slowest process in the particle dynamics (for $\alpha > \beta$ and $\beta < 1/2$), the system is in a high-density (HD) phase with steady-state properties given by
\begin{equation} \label{HD}
J_{HD}= \beta (1- \beta), \quad \rho_{bulk,HD}=1-\beta.
\end{equation}
For $\alpha > 1/2$ and $\beta >1/2$ the dynamics of the system is determined by processes in the bulk, and we have a maximal-current (MC) phase with
\begin{equation} \label{MC}
J_{MC}= \frac{1}{4}, \quad \rho_{bulk,MC}=\frac{1}{2}.
\end{equation}

Since every channel can be found in one of three stationary phases, there are nine possible phases in the two-lane ASEP with narrow entrances, assuming that the dynamics in the channels is independent of each other. However, some of the phases might not exist. Because of the coupling at the boundaries of the channels some of the phases could be asymmetric. To investigate this possibility let us consider separately symmetric and asymmetric cases.

\subsubsection{Symmetric Phases}

In symmetric phases all  properties in both channels are identical,
\begin{equation}
J_{1}=J_{2},  \quad \rho_{1}=\rho_{2}, \quad \alpha_{1}=\alpha_{2},
\end{equation}
where $J_{i}$ and $\rho_{i}$ are the stationary-state particle current and the bulk density in the channel $i$ (with $i=0$, 1). Because of the symmetry, the probabilities to find particles in both channels are related via
\begin{equation}
p_{i}=m_{L-i+1}.
\end{equation}

In MC phase the stationary current is equal to $J_{1}=J_{2}=1/4$. From the expression for the current at the exit of the channel 2,
\begin{equation}
J_{2}=\beta m_{1}=1/4,
\end{equation}  
we derive $m_{1}=\frac{1}{4\beta}$. Substituting this result into Eq. (\ref{eq_1}) yields the value for the effective entrance rate,
\begin{equation} 
\alpha_{1}=\alpha(1- \frac{1}{4 \beta}).
\end{equation}
Then the conditions for the existence of this phase ($\alpha_{1} > 1/2$ and $\beta > 1/2$) can be written as
\begin{equation}
\alpha > \frac{2 \beta}{4 \beta -1}.
\end{equation}

In LD phase the current in both channels is equal to $J_{1}=J_{2}=\alpha_{1}(1-\alpha_{1})$. The exit current from the lane 2 is equal to
\begin{equation}
J_{2}=\beta m_{1}=\alpha_{1}(1-\alpha_{1}),
\end{equation}  
and we obtain $m_{1}=\frac{\alpha_{1}(1-\alpha_{1})}{\beta}$. After substituting this result into Eq. (\ref{eq_1}), it can be shown that
\begin{equation} 
\alpha_{1}=\frac{\alpha+\beta - \sqrt{(\alpha+\beta)^{2}-4 \alpha^{2}\beta}}{2 \alpha}.
\end{equation}
Because in the low-density phase $\alpha_{1}< \beta$ and $\alpha_{1} < 1/2$, one can show that the condition for existence of this phase is the following, 
\begin{equation}
\alpha < \frac{2 \beta}{4 \beta -1}.
\end{equation}

Another possible symmetric phase is HD phase where $J_{1}=J_{2}=\beta (1- \beta)$. Because the exit current in the channel 2 is $J_{2}=\beta m_{1}$, we  can show that $m_{1}=1-\beta$. The existence of this phase is specified by the condition $\alpha_{1}=\alpha(1-m_{1}) > \beta$, and it can take place only for $\alpha > 1$. This suggests that the HD phase cannot be found in two-lane ASEP with narrow entrances.

\subsubsection{Asymmetric Phases} \label{asymm}

In asymmetric phases the currents and densities in each channel are generally different,
\begin{equation} 
J_{1}\ne J_{2}, \quad \alpha_{1}\ne \alpha_{2}.
\end{equation} 
First, let us consider the possibility of phases when one of the lanes supports the maximal-current phase. For convenience, assume that the channel 1 is in the maximal-current state. The MC/HD phase is specified by
\begin{equation}
\left \{ \begin{array}{cc} 
                         \alpha_{1} > 1/2   & \mbox {  and } \beta > 1/2, \nonumber \\
                         \alpha_{2} > \beta & \mbox { and } \beta < 1/2.
\end{array} \right.
\end{equation}
However, these conditions contradict each other, and we conclude that MC/HD phase does not exist.

Similarly, the conditions of existence for MC/LD phase can be written as
\begin{equation}
\left \{ \begin{array}{cc} 
                            \alpha_{1} > 1/2   & \mbox {  and } \beta > 1/2, \nonumber \\
                            \alpha_{2} < \beta & \mbox { and } \alpha_{2} < 1/2.
\end{array} \right.
\end{equation}
The stationary currents in the system are given by
\begin{equation}
J_{1}=\beta p_{L}=1/4, \quad J_{2}=\beta m_{1} = \alpha_{2} ( 1- \alpha_{2}),
\end{equation}
which leads to the following expressions,
\begin{equation}
p_{L}=\frac{1}{4 \beta}, \quad m_{1}=\frac{\alpha (1-\frac{1}{4\beta})\left[1-(1-\frac{1}{4\beta})\right]}{\beta}.
\end{equation}
Because $\alpha_{1}=\alpha(1-m_{1}) > 1/2$, it requires that
\begin{equation}
\frac{\alpha^{2}}{\beta} \left[ \frac{1}{2 \alpha}- (1-\frac{1}{4\beta}) \right] >1.
\end{equation}
However, this expression cannot be satisfied for $0 < \alpha \le 1$ and $0 < \beta \le 1$, and it leads us to conclusion that the MC/LD phase also cannot exist in the two-lane ASEP with narrow entrances.

The situation is different for the HD/LD phase  which is defined by the following expressions,
\begin{equation}\label{HD/LD}
\left \{ \begin{array}{cc}
          \alpha_{1} > \beta   & \mbox {  and } \beta < 1/2, \nonumber \\
           \alpha_{2} < \beta & \mbox { and } \alpha_{2} < 1/2.
\end{array} \right.
\end{equation}
The stationary current in this phase can be written as
\begin{equation}
J_{1}=\beta p_{L}=\beta(1-\beta), \quad J_{2}=\beta m_{1} = \alpha_{2} ( 1- \alpha_{2}),
\end{equation}
and these expressions lead to
\begin{equation}
\alpha_{1}=\alpha(1-\alpha +\alpha^{2} \beta), \quad \alpha_{2}=\alpha \beta.
\end{equation}
Combining these equations with the conditions (\ref{HD/LD},) we derive the following condition for the existence of  this phase,
\begin{equation}
\beta < \frac{\alpha}{1+\alpha+ \alpha^{2}}.
\end{equation}

Another possible asymmetric phase is the LD/LD phase, where both channels are in the low-density state but with different particle currents and bulk densities. Formally, this phase can be described by 
\begin{equation}\label{LD/LD}
\left \{ \begin{array}{cc}
         \alpha_{1} < \beta   & \mbox {  and  } \alpha_{1} < 1/2, \nonumber \\
         \alpha_{2} < \beta & \mbox { and  } \alpha_{2} < 1/2. 
\end{array} \right.
\end{equation}
The stationary currents in both channels are given by
\begin{equation}
J_{1}=\beta p_{L}=\alpha_{1}(1-\alpha_{1}), \quad J_{2}=\beta m_{1} = \alpha_{2} ( 1- \alpha_{2}),
\end{equation}
and from these expressions we derive
\begin{equation}\label{alpha}
\alpha_{1}=\alpha \left[1-\frac{\alpha_{2}(1-\alpha_{2})}{\beta}\right], \quad \alpha_{2}=\alpha \left[1- \frac{\alpha_{1}(1-\alpha_{1})}{\beta}\right].
\end{equation}
To simplify calculations, assume that $\alpha_{1}>\alpha_{2}$ and introduce two auxiliary functions,
\begin{equation}
S=\alpha_{1}+\alpha_{2}, \mbox {  and  } D=\alpha_{1}-\alpha_{2}.
\end{equation}
From Eqs. (\ref{alpha}) one can show that
\begin{equation}
D=\frac{\alpha}{\beta} (\alpha_{1}-\alpha_{1}^{2}-\alpha_{2}+\alpha_{2}^{2})=\frac{\alpha}{\beta} D (1-S).
\end{equation}
Since in the asymmetric phase $D \ne 0$ it leads to
\begin{equation}\label{eq_S}
S=1-\beta/\alpha.
\end{equation}
Similarly, from Eqs. (\ref{alpha}) we obtain 
\begin{equation}
S=2\alpha - \frac{\alpha}{\beta}\left[ S - (S^{2}+D^{2})/2 \right].
\end{equation}
Then, using the result (\ref{eq_S}), it can be shown that
\begin{equation}\label{eq_D}
D=\sqrt{1-4\beta +2 \left(\frac{\beta}{\alpha}\right) - 3 \left(\frac{\beta}{\alpha}\right)^{2}}.
\end{equation}
The effective entrance rates can be easily calculated in this case from Eqs. (\ref{eq_S}) and (\ref{eq_D}),
\begin{eqnarray}
\alpha_{1}=(S+D)/2&=\frac{1}{2} \left[ 1- \left(\frac{\beta}{\alpha}\right) + \sqrt{1-4\beta +2 \left(\frac{\beta}{\alpha}\right) - 3 \left(\frac{\beta}{\alpha}\right)^{2}} \right], \nonumber \\
\alpha_{2}=(S-D)/2&=\frac{1}{2} \left[ 1- \left(\frac{\beta}{\alpha}\right) - \sqrt{1-4\beta +2 \left(\frac{\beta}{\alpha}\right) - 3 \left(\frac{\beta}{\alpha}\right)^{2}} \right].
\end{eqnarray}
This phase can only exist if the expression inside of the square root is positive. This requirement, in addition with Eqs. (\ref{LD/LD}), determines the conditions for existence of the LD/LD asymmetric phase in the mean-field approximation,
\begin{equation}
\frac{\alpha}{1+\alpha+ \alpha^{2}} < \beta < \frac{\alpha(1-2\alpha) + 2\alpha\sqrt{\alpha^{2}-\alpha+1}}{3}.
\end{equation}

The resulting phase diagram, obtained via the mean-field calculations, is shown in Fig. 2. Similarly to the single-lane two-species ASEP where the symmetry breaking was first observed \cite{evans95,EFGM,arndt98,clincy01,erickson05}, the region for existence of the LD/LD asymmetric phase is very small. But there is  a MC phase in our system, not observed in the single-lane two-species model. In addition, the mean-field approach predicts a first-order phase transition between HD/LD and LD/LD asymmetric phases, and continuous phase transformations for boundaries between asymmetric LD/LD and symmetric LD phases, and between symmetric LD and MC phases.

\begin{figure}[h]
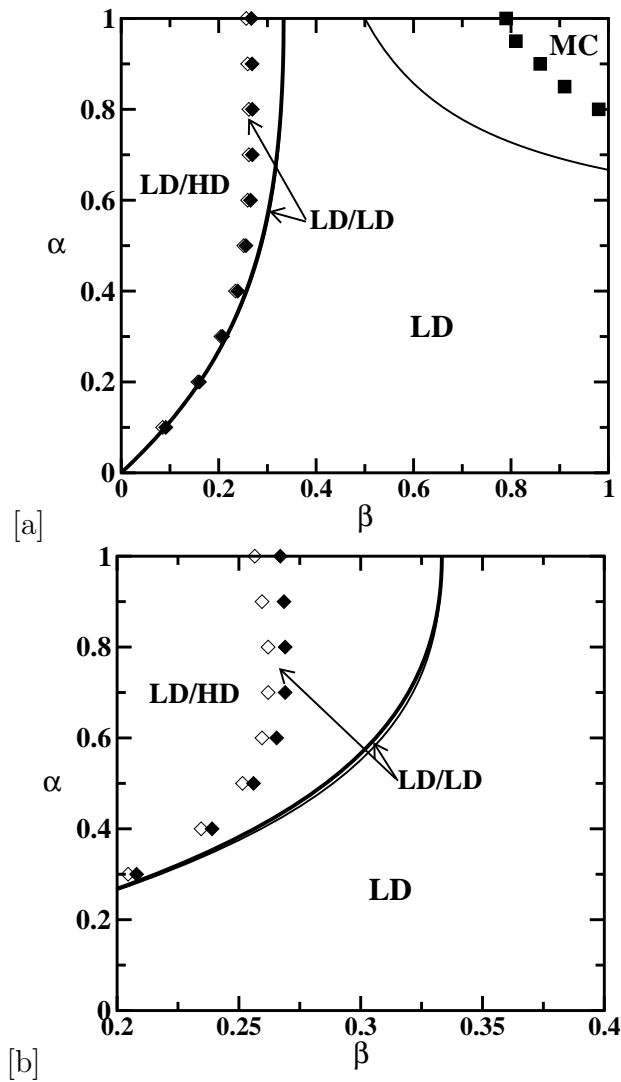
 
\centering
[a]\includegraphics[scale=0.4, clip=true]{Fig2a.eps}
[b]\includegraphics[scale=0.4, clip=true]{Fig2b.eps}
\caption{Phase diagram for a two-channel ASEP with narrow entrances. Lines are predictions from the mean-field theory. Thick solid lines represent first-order phase transitions, while thin solid lines correspond to continuous transitions. Symbols are from Monte Carlo computer simulations for the channels with $L=1000$. Filled squares describe the continuous  transitions between symmetric LD and MC phases. Filled diamonds describe  the continuous  transitions between symmetric LD and asymmetric LD/LD phases, while unfilled diamonds are at the position of first-order phase transition between HD/LD and LD/LD asymmetric phases. a) Phase diagram for all parameter space; b) phase diagram for $0.2 < \beta < 0.4$.}
\label{phase_diag_L=1000}
\end{figure}

\section{Monte-Carlo Simulations and Discussions}

Our theoretical approach neglects correlations between particles, and it is necessary to test theoretical predictions  with  computer simulations. We performed extensive Monte Carlo computer simulations to analyze two-channel two-species ASEP with narrow entrances. Theoretical calculations assume that the system is in the thermodynamic limit, i.e., $L \rightarrow \infty$. However, in computer simulations only the processes with finite size channels can be tested. The systems with lattice sizes ranging from $L=100$ up to $L=12,000$ have been used in our simulations, but most computations have been done for $L=1000$. To accelerate the simulations we implemented the so-called BKL algorithm \cite{BKL}. This method skips uneventful Monte Carlo steps with simultaneous correction in the calculation of the time. Number of effective Monte Carlo steps per lattice site  in our simulations was typically between $2 \times 10^{7}$ and $5 \times 10^{8}$. Usually, first five percents of steps have been omitted from the calculation because we are only interested in the stationary state of the system.

Since the phase diagram predicted by the mean-field theoretical approach for two-channel ASEP with narrow entrances is very complex,  it is not enough to measure in the simulations only the average densities and the currents per channel. It was proposed earlier \cite{arndt98,clincy01} to utilize also density distribution functions that measure the frequency of simultaneously observing average densities for different species.  It was shown that this method was powerful enough to map precisely the phase diagram of the single-lane two-species ASEP. In our system the density distribution function depends on the average densities in the first and the second channels determined at the same time. To construct the corresponding distributions these densities have been measured after every $L/10$ Monte Carlo steps. The resulting histograms are shown in Fig. 3 for the fixed value of $\alpha=0.9$, where two asymmetric and two symmetric phases can be clearly seen. 

\begin{figure}[h] 
\centering
[a]\includegraphics[scale=0.3,clip=true]{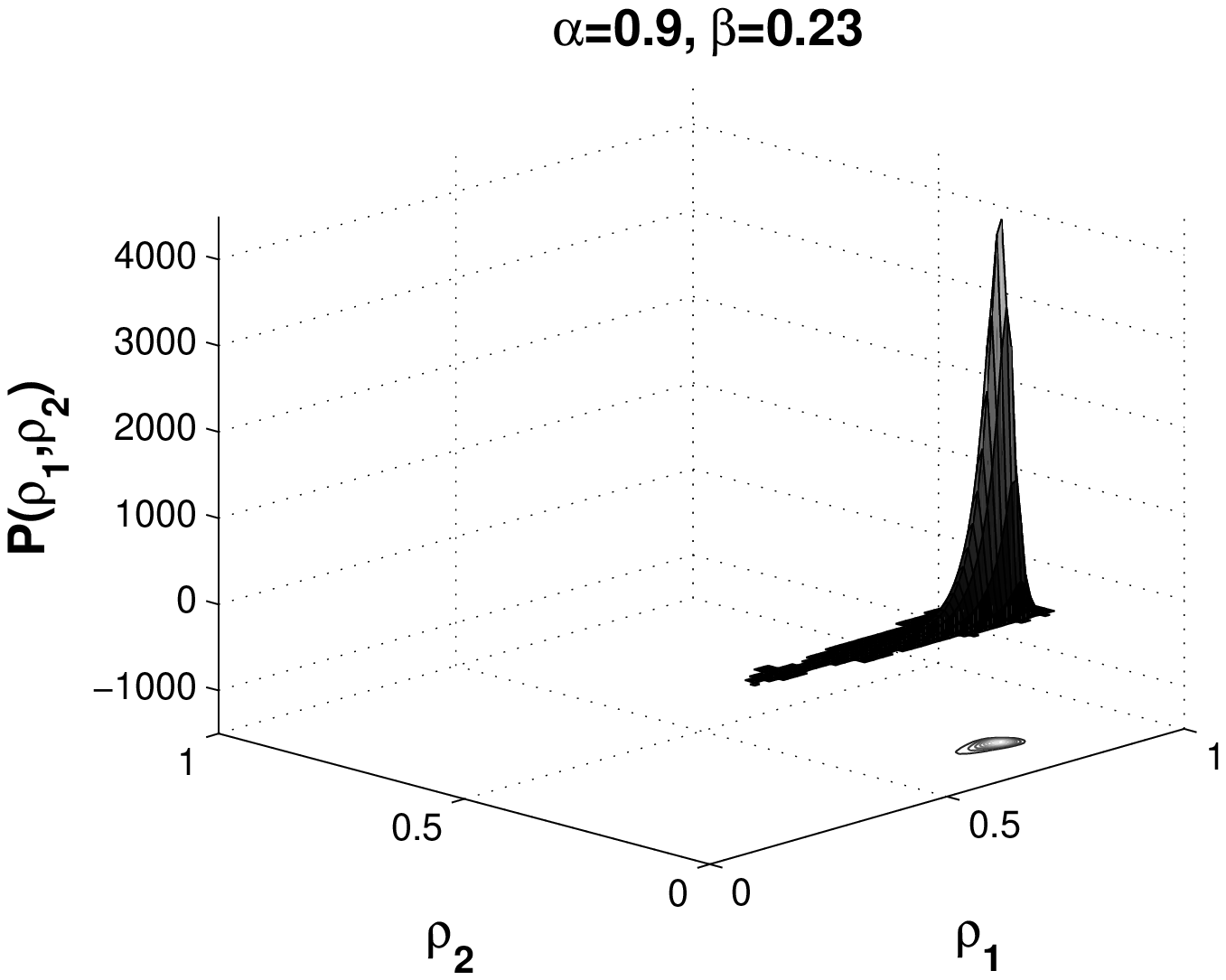}
[b]\includegraphics[scale=0.3,clip=true]{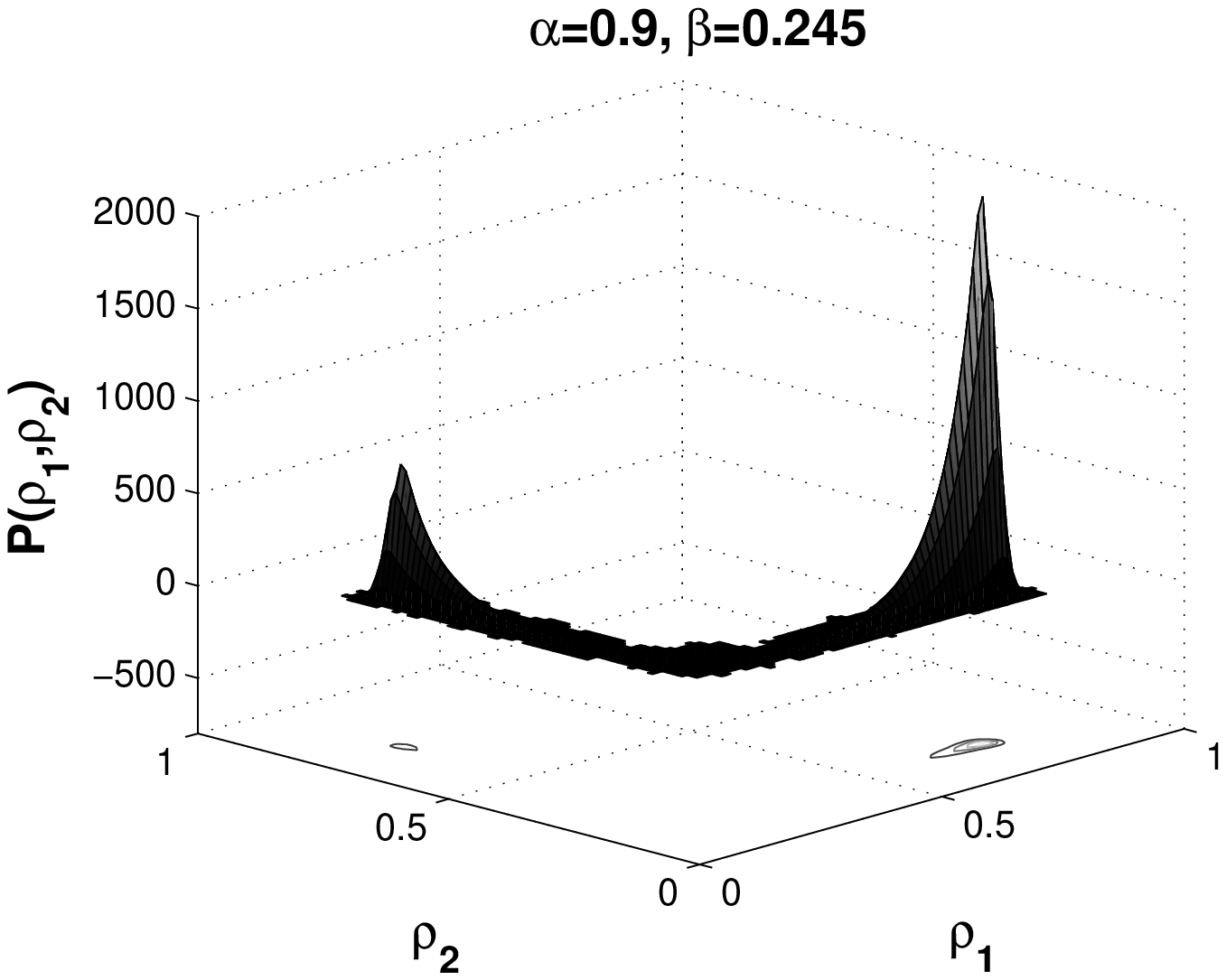}
[c]\includegraphics[scale=0.3,clip=true]{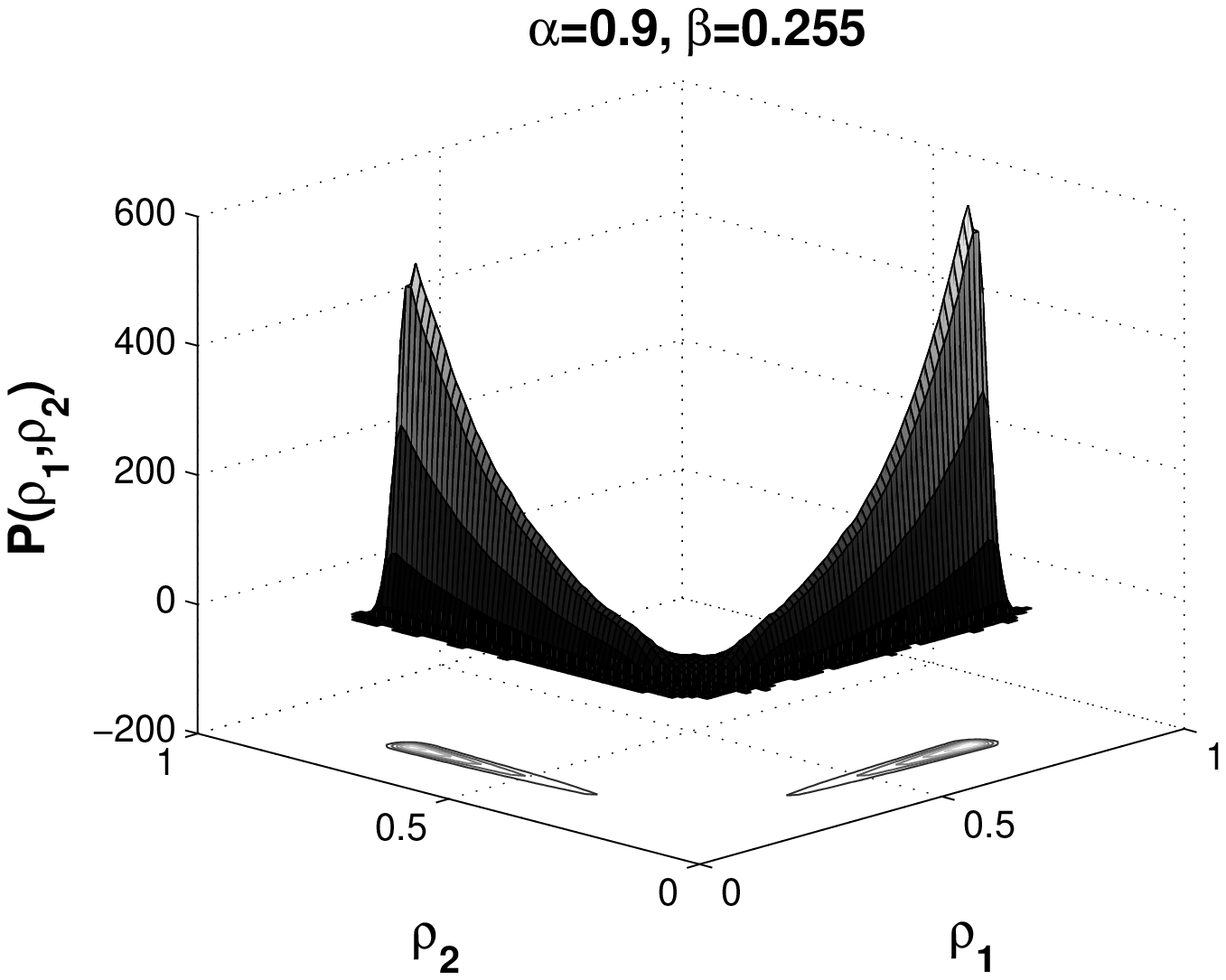}
[d]\includegraphics[scale=0.3,clip=true]{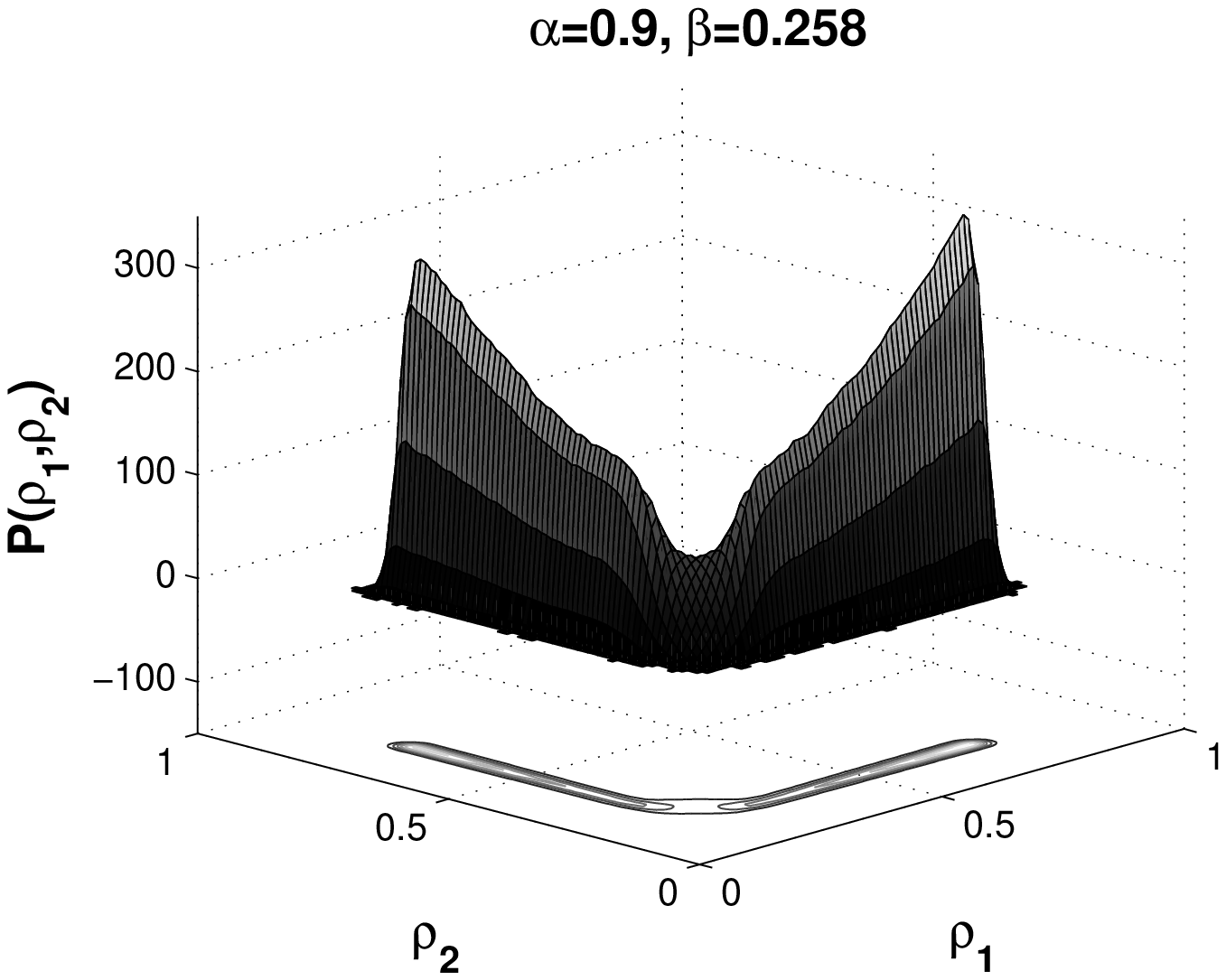}
[e]\includegraphics[scale=0.3,clip=true]{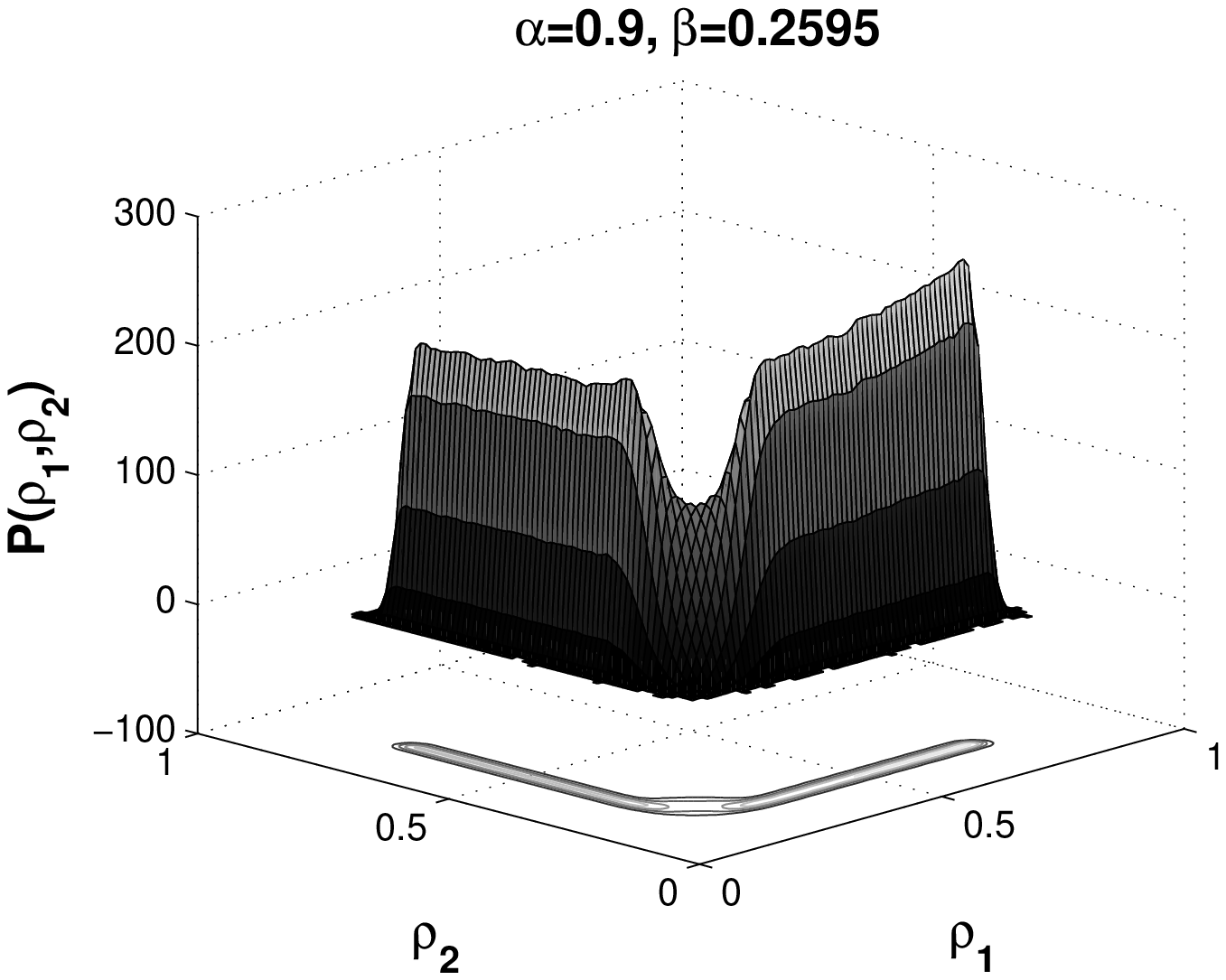}
[f]\includegraphics[scale=0.3,clip=true]{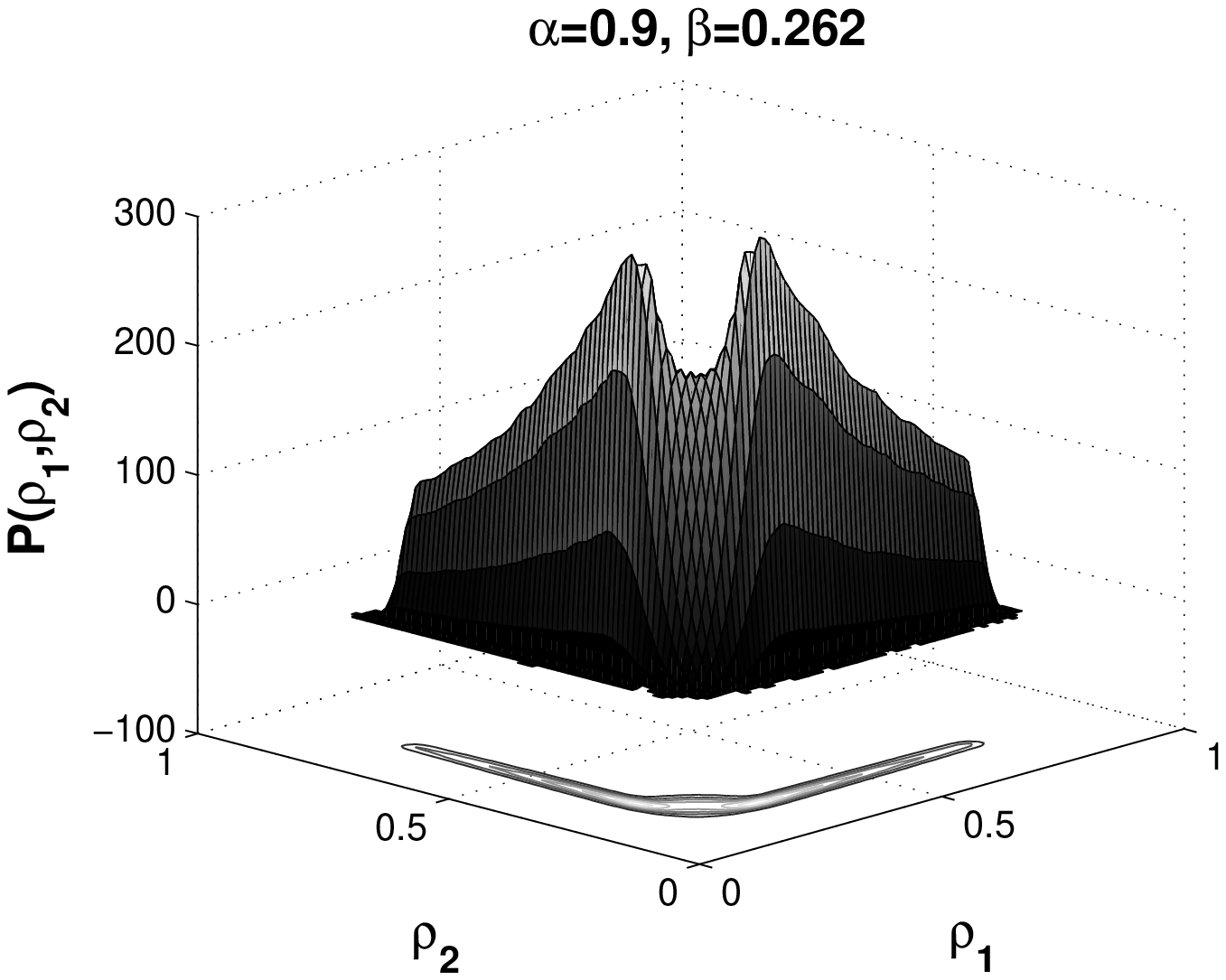}
[g]\includegraphics[scale=0.3,clip=true]{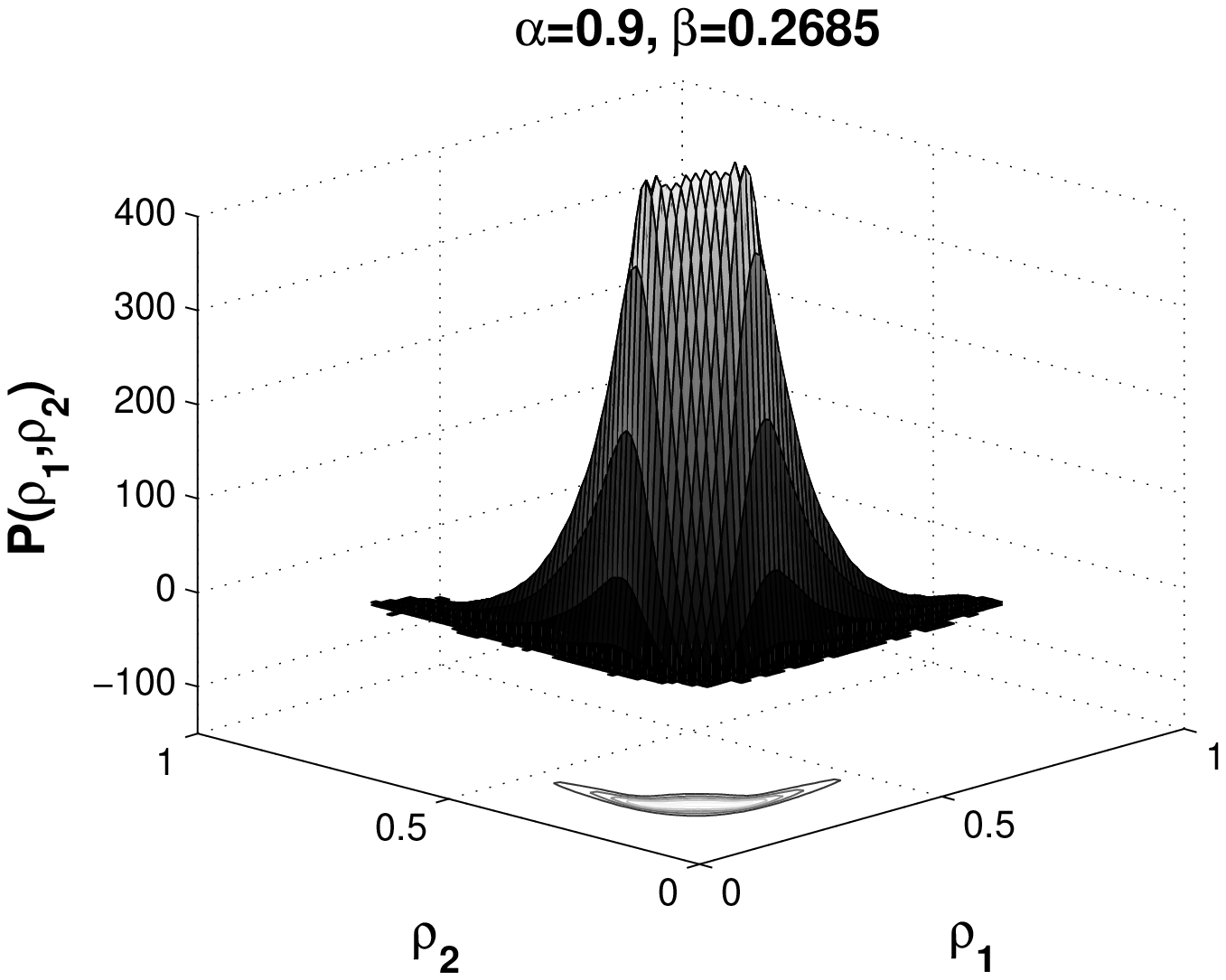}
[h]\includegraphics[scale=0.3,clip=true]{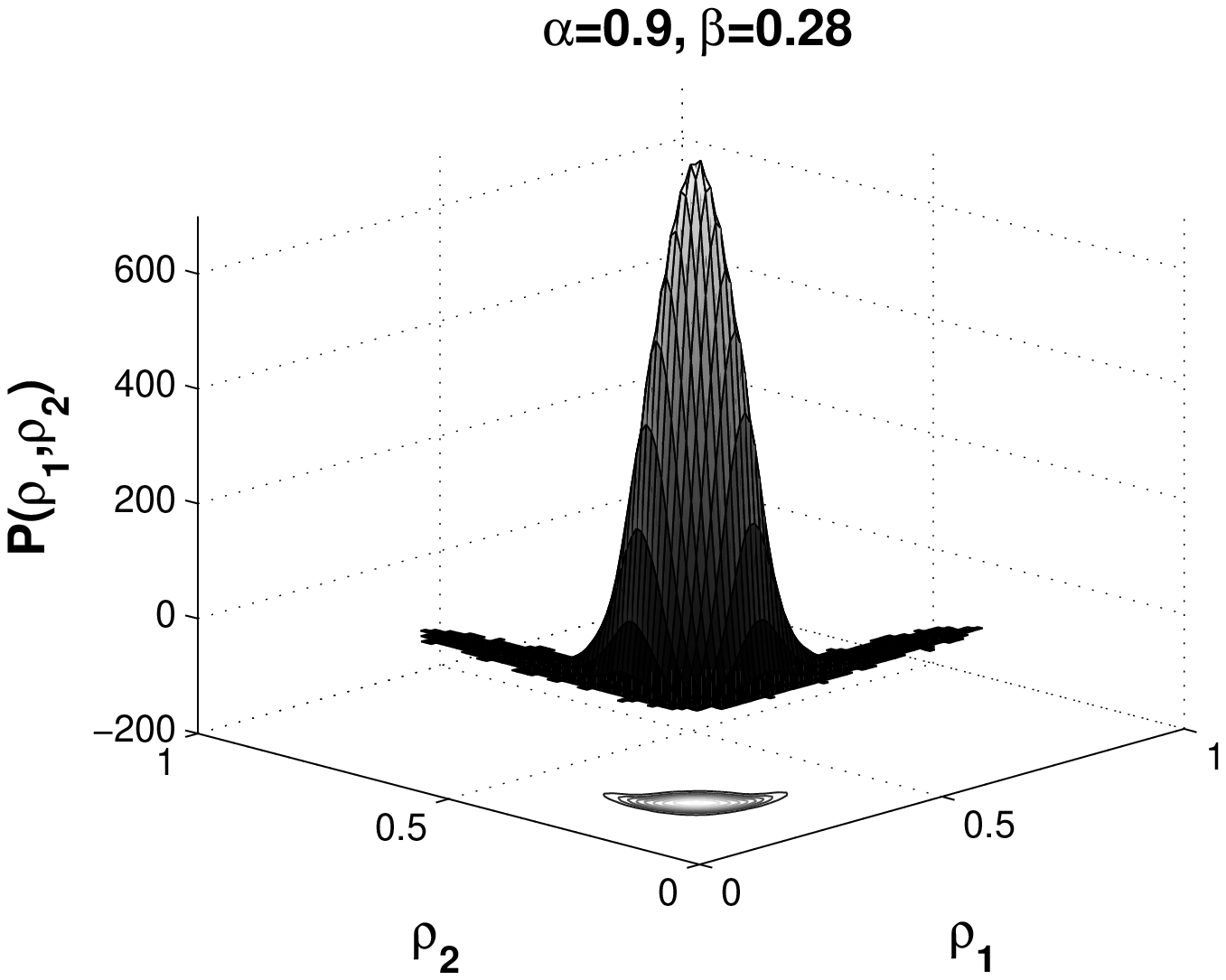}
[i]\includegraphics[scale=0.3,clip=true]{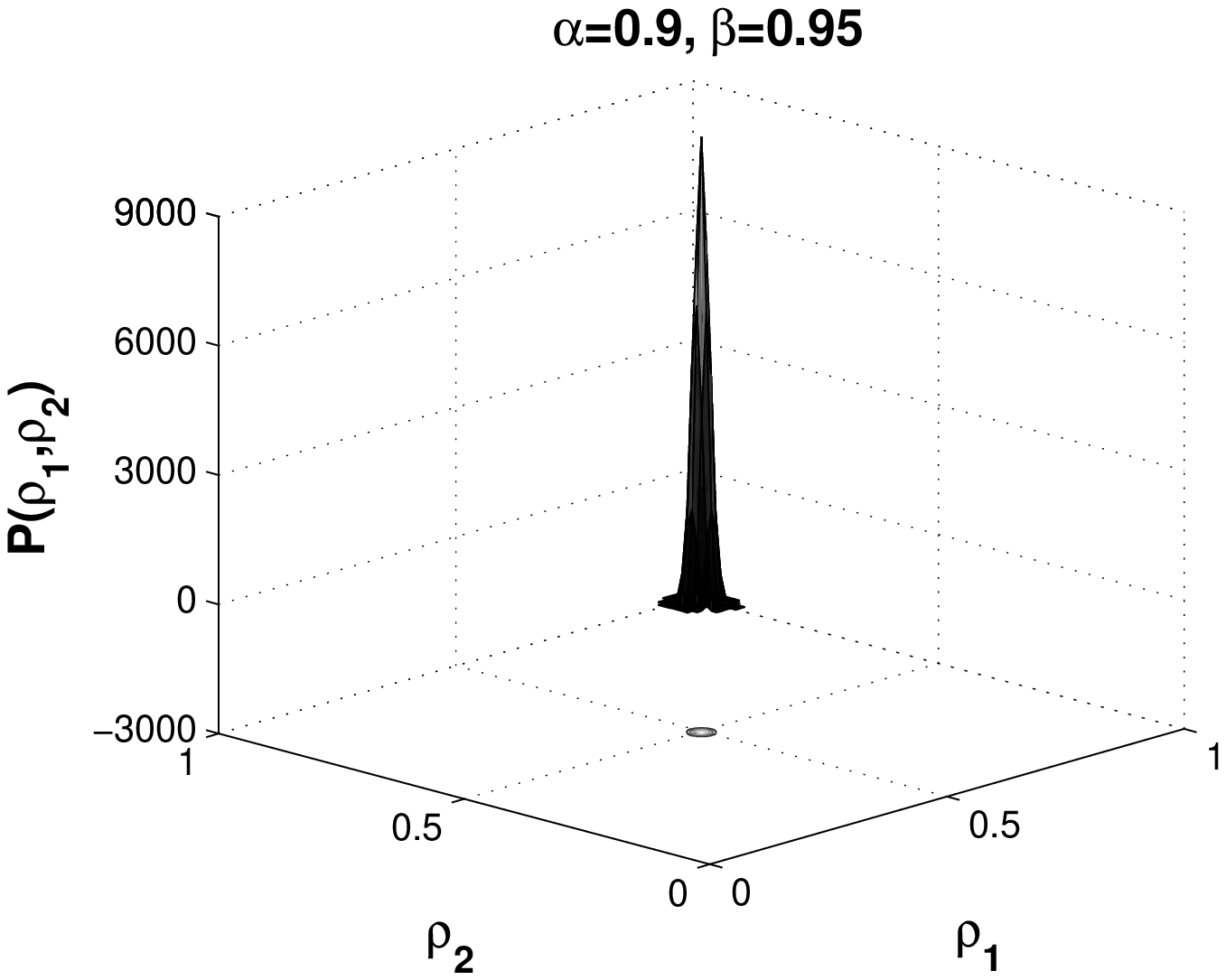}
\caption{3D plots of the probability density $P(\rho_{1},\rho_{2})$ as a function of the average particle densities in both channels for $L=1000$ and $\alpha=0.9$. Contour plots are shown projected onto $\rho_{1}-\rho_{2}$ plane. (a) asymmetric HD/LD phase at $\beta=0.23$; (b) asymmetric HD/LD phase at $\beta=0.245$; (c) asymmetric HD/LD phase at $\beta=0.255$; (d) asymmetric HD/LD phase at $\beta=0.258$; (e) phase coexistence between asymmetric HD/LD and LD/LD phases at $\beta=0.2595$; (f) asymmetric LD/LD phase at $\beta=0.262$; (g) phase coexistence between asymmetric LD/LD and symmetric LD phases at $\beta=0.2685$; (h) symmetric LD phase at $\beta=0.28$; (i) symmetric MC phase at $\beta=0.95$.   }
\end{figure}

In addition, to determine exactly the position of the boundary between symmetric LD and MC phases the derivative of the average current as a function of the exit rate $\beta$ has been computed for different fixed values of the entrance rate $\alpha$. In the MC phase the current reaches a constant maximum value, and we associate the phase boundary with the value of $\beta$ when the derivative reaches zero or starts to fluctuate around zero. This procedure is very robust and it is illustrated in Fig. 4.

\begin{figure}[h]
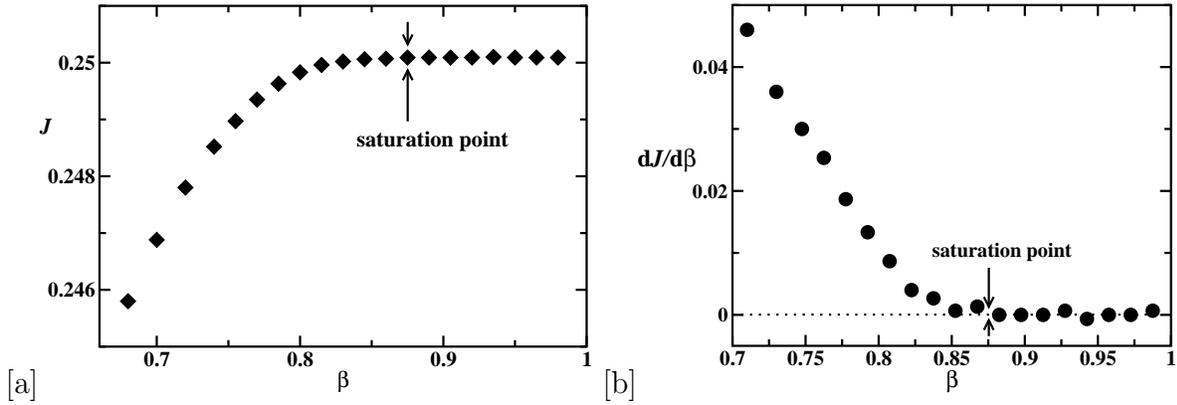
 
\centering
[a]\includegraphics[scale=0.3, clip=true]{Fig4a.eps}
[b]\includegraphics[scale=0.3, clip=true]{Fig4b.eps}
\caption{ Computer simulations results for (a) current and (b) derivative of the current as a function of the exit rate $\beta$ for $L=4000$ and $\alpha=0.9$. Arrows show the current saturation point that we associate with the phase boundary between symmetric LD and MC phases.}
\end{figure}

One of the most controversial issues in studies of symmetry breaking in the single-lane two-species ASEP is the question of existence of the asymmetric LD/LD phase \cite{arndt98,clincy01,erickson05}. Similar situation is also observed in the two-lane two-species ASEP with narrow entrances. Our mean-field theory predicts that the asymmetric LD/LD phase is real, although the range of existence is quite small - see Fig. 2. Monte Carlo simulations performed for the channels of size $L=1000$ also suggest that this phase is a stable stationary-state regime. However, it was shown earlier that there are very strong size-scaling effects in ASEPs with symmetry breaking \cite{erickson05}. In order to study  this size-scaling dependency in our model  we carried out computer simulations for the systems with different sizes (up to $L=12,000$). Phase boundaries have been computed in Monte Carlo simulations for different system sizes by utilizing the density-distribution functions and current derivatives method, as described above, and the results are presented in Fig. 5. It is shown in Fig. 5a that the boundary between the the asymmetric HD/LD and LD/LD phases does not depend on the system size, while the region of existence for the asymmetric LD/LD phase shrinks constantly with increasing $L$ without reaching a saturation. This suggests that in the thermodynamic limit ($L \rightarrow \infty$) the asymmetric LD/LD phase does not exists. In addition, the results presented in Fig. 5b indicate that the symmetric MC phase also depends on the system size. With increasing $L$ the boundary between symmetric LD and MC phases shifts closer to the theoretical predictions, but it is not clear if it will reach it in the thermodynamic limit. 

\begin{figure}[h]
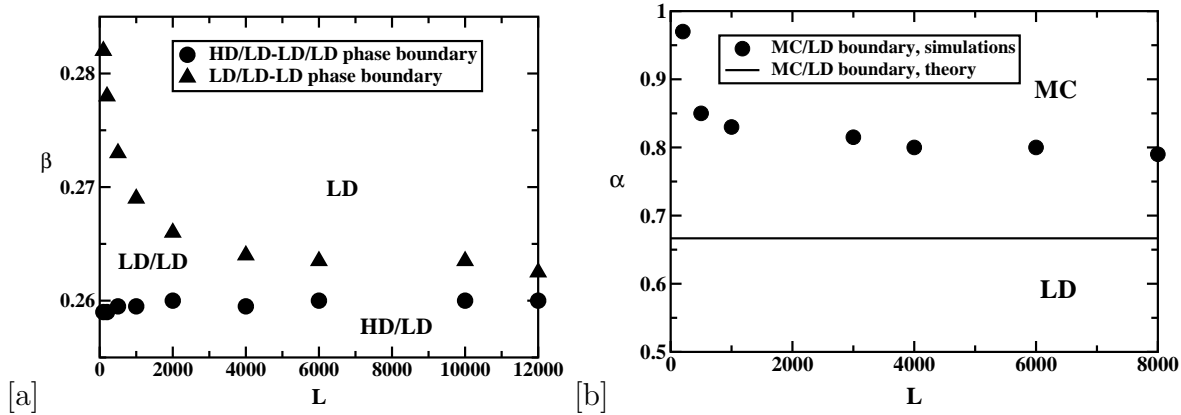
 
\centering
[a]\includegraphics[scale=0.3, clip=true]{Fig5a.eps}
[b]\includegraphics[scale=0.3, clip=true]{Fig5b.eps}
\caption{ Phase boundaries as a function of the channel size $L$. (a) Computer simulations results for phase boundaries between the asymmetric HD/LD, the asymmetric LD/LD and the symmetric LD phases for $\alpha=0.9$; (b) Computer simulations results for phase boundary between symmetric MC and LD phases for $\beta=1$. Solid line corresponds to the phase boundary calculated in the mean-field  approximation.}
\end{figure}

The phase diagram computed from the Monte Carlo simulations is shown in Fig. 2, while bulk densities and particle currents are presented in Fig. 6. By comparing the results of the numerical simulations with the theoretical predictions  we conclude that the mean-field approach agrees with simulations only for small values of entrance and exit rates. For large values of $\alpha$ and $\beta$ theoretical calculations, although qualitatively correct, deviate significantly from Monte Carlo computer simulations. In addition, the mean-field approach predicts two asymmetric (HD/LD and LD/LD) and two symmetric (LD and MC) phases, however our size-scaling analysis of computer simulations (see Fig. 5) shows that asymmetric LD/LD phase does not exist in the thermodynamic limit. It is also interesting to note that two-channel two-species ASEP with narrow entrances support the maximal-current phase, in contrast to the single-lane two-species ASEP \cite{evans95,EFGM,arndt98,clincy01,erickson05}.

The fact that the mean-field approach cannot fully describe phase diagram and stationary-state properties of the two-lane two-species ASEP with narrow entrances indicates that correlations are important in the dynamics of this system, similarly to the single-lane two-species ASEP \cite{evans95,EFGM,arndt98,clincy01,erickson05}. However, in our system two channels do not interact with each other except at the entrance/exit lattice sites, and it is very surprising that this local interaction strongly influences the macroscopic particle dynamics in the system. This observation supports the idea that symmetry breaking in ASEPs might occur only in the systems with the effective boundary defects \cite{popkov04}.

\begin{figure}[h]
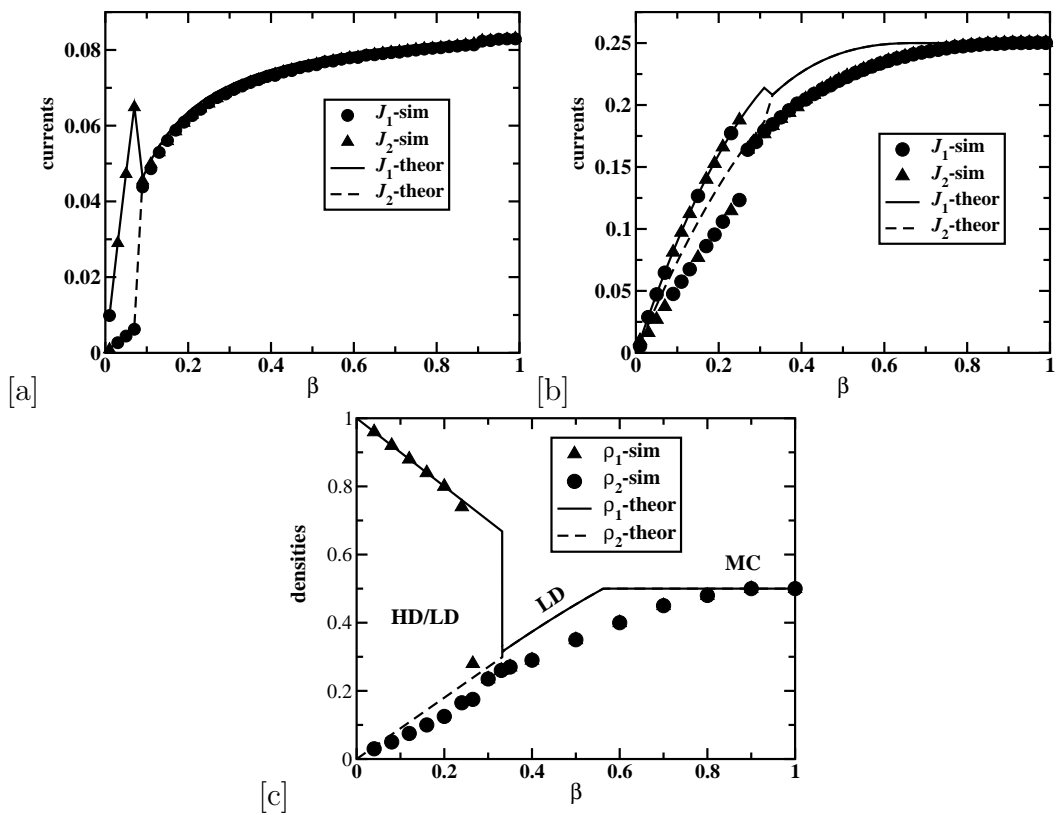
 
\centering
[a]\includegraphics[scale=0.3, clip=true]{Fig6a.eps}
[b]\includegraphics[scale=0.3, clip=true]{Fig6b.eps}
[c]\includegraphics[scale=0.3, clip=true]{Fig6c.eps}
\caption{ Stationary-state properties as a function of $\beta$. Lines are predictions from the mean-field theory, symbols are from Monte Carlo computer simulations for $L=1000$. (a) currents in both channels for $\alpha=0.1$; (b) currents in both channels for $\alpha=0.8$; (c) bulk densities in both channels for $\alpha=0.9$.}
\end{figure}

\section{Summary and Conclusions} \label{summary}

We have investigated two-channel  ASEP with two types of particles moving in opposite directions along the channels with narrow entrances via the mean-field theory, that neglects correlations in the system, and extensive Monte Carlo computer simulations. Our theoretical analysis indicates that there are four stationary-state phases in the system, with two of them exhibiting symmetry breaking in  dynamic properties. However, computer simulations only partially support the mean-field picture. Symmetry breaking is found in HD/LD phase, and it is shown that the existence of asymmetric LD/LD phase is a size-scaling effect: the phase disappears in the thermodynamic limit. In addition, mean-field predictions for particle currents, bulk densities and phase boundaries agree with computer simulations quantitatively only for the low entrance and exit rates, while in other regions of the parameter space the agreement is mostly qualitative.

In our system particles from the different channels interact only at the entrance sites. However, surprisingly, this local interaction still leads to symmetry breaking and long-range correlations in the system. We conclude that these phenomena can be viewed as a result of putting an effective impurity on the boundary of the lattice. It is important to develop a better theoretical description of these processes that can explain the mechanisms of symmetry breaking in more detail. It is reasonable to suggest that the domain-wall method \cite{kolomeisky98},  successful in the determination of mechanisms and properties of non-equilibrium processes in ASEPs, is the most promising approach for future investigations.   

\ack

The support  from the Welch Foundation (under Grant No. C-1559), Hammill Research Innovation Award and from the US National Science Foundation through the grant CHE-0237105 is gratefully acknowledged.

\end{document}